 \documentclass[12pt,preprint]{aastex}

\usepackage{graphicx}
\usepackage{natbib}
	\newcommand{\citeN}[1]{\citeauthor{#1} (\citeyear{#1})}
	\newcommand{\citeNP}[1]{\citeauthor{#1} \citeyear{#1}}
	\newcommand{\citeyearNP}[1]{\citeyear{#1}}
	\newcommand{\citeANP}[1]{\citeauthor{#1}}

   \shortauthors{Socas-Navarro \& S\'anchez Almeida}
   \shorttitle{Magnetic properties of photospheric regions having 
	very low magnetic flux}

\begin{document}

   \title{
   Magnetic properties of photospheric regions with
	very low magnetic flux
%   \new{Determination of magnetic properties in photospheric 
%	regions having very low magnetic flux}
   }

   \author{H. Socas-Navarro}
   	\affil{High Altitude Observatory, NCAR\thanks{The National Center
	for Atmospheric Research (NCAR) is sponsored by the National Science
	Foundation.}, Boulder, CO 80307-3000, USA}
	\email{navarro@hao.ucar.edu}
	
   	 \and
   	  
\author{J. S\'anchez Almeida}
	\affil{Instituto de Astrof\' \i sica de Canarias, E-38200, La Laguna,
	Tenerife, Spain}
	\email{jos@ll.iac.es}

%   \offprints{H. Socas-Navarro}

   \date{}%

   \begin{abstract}
	The magnetic properties of the quiet Sun are investigated using
	a novel inversion code, FATIMA, based on the Principal
	Component Analysis of the observed Stokes profiles. The stability
	and relatively low noise sensitivity of this inversion procedure
	allows for the systematic inversion of large data sets with very weak
	polarization signal. Its application to quiet Sun observations of
	network and internetwork regions reveals that a significant fraction
	of the quiet Sun contains kilogauss fields (usually with very
	small filling factors) and confirms that the pixels with weak
	polarization account for most of the magnetic flux.
	Mixed polarities in the resolution element
	are also found to occur more likely as the
	polarization weakens.	
   \end{abstract}
   
      \keywords{line: profiles -- methods: numerical --
                Sun: atmosphere -- Sun: magnetic fields --
                Sun: photosphere}

%
%________________________________________________________________

%

%\section{}
%\new{This corresponds to new or modified text}
%
%\old{ Old wording retained for quick comparison. To be
%	deleted in the next version}.
%
%\task{Work that remains to be done}

\section{Introduction}
\label{intro}

The study of the quiet Sun magnetism is 
an open and challenging problem,
particularly in the low magnetic flux limit ($\sim$ 10$^{17}$ Mx
and below). From the observational point of view, two important
difficulties must be overcome. First, the polarization signals are
weak and therefore close to the noise level. This complicates the analysis of the observations with
the currently available diagnostic tools. They employ non-linear least-squares
fitting algorithms, which are usually very unstable in this
regime. 
A second difficulty comes from the fact that the magnetic structures
cannot be spatially resolved by present-day instrumentation. 
The measurement of magnetic properties has to be based on modeling.
However, without a reliable physical model for the magnetic elements, 
the inference of atmospheric
properties is uncertain. Any sensible
diagnostic technique must
necessarily rely on a proper understanding of the basic properties of the
atmosphere (see, e.g., the discussion in \citeNP{soc01}).
The complication arises because both theoretical and
observational arguments indicate that the magnetic field in the quiet solar
photosphere may be intermittent with length scales much smaller than the mean
free path of the photons (e.g., \citeNP{san96}; \citeNP{san98c}; 
\citeNP{cat99a}). In this
situation the description of the basic properties of the
atmosphere is not simple. The radiative transfer becomes non-trivial and absorption
and emission  must be properly averaged at each spatial point
to synthesize spectra (\citeNP{san96}).
 The acronym MISMA
(MIcro-Structured Magnetic Atmosphere) denotes an atmosphere with such
small-scale magnetic inhomogeneities.

The problem of inverting Stokes profiles formed in MISMAs was addressed by 
\citeN{san97b}, who developed an inversion code (IC) using a standard
non-linear least-squares algorithm\footnote{The term {\em inversion code} stands for
	any procedure aimed at retrieving information on the physical conditions in the solar atmosphere
	by fitting observed quantities. The Stokes profiles describe
	the polarization of a spectral line, hence {\em inverting Stokes profiles}
	means inferring physical properties by reproducing the observed line polarization.}.
This code was applied by 
\citeANP{san00} (\citeyearNP{san00}, hereafter SAL) to quiet Sun
observations from the Advanced Stokes Polarimeter (\citeNP{elm92}).
 Since the polarization signal was too weak in most of
the points of the observed map, SAL inverted only 15\% of 
surface,
where the polarization was above some arbitrary
threshold. As a result, they obtained 5200 different model MISMAs that are
capable of fitting the observations, even the most strikingly asymmetric
ones, within a single framework.

Our work generalizes the study of SAL applying a novel inversion
procedure to a larger fraction of the same observed region. 
By doing so, we
extend the original
results to weaker and therefore more numerous signals, thus improving the
statistical significance of the conclusions. We also 
demonstrate the feasibility of the new technique to systematically
retrieve information on the weak and very asymmetric line polarization
produced by the quiet Sun, for which no other technique is available
at the moment. 
The inversion
technique is denoted by FATIMA 
(Fast Analysis Technique for the Inversion of Magnetic Atmospheres), and
a  brief is provided in \S \ref{fatima}. 
We begin this work by summarizing the observational material (\S \ref{observations}). 
Then 
the specificities of the general inversion technique  
applied to our particular data set are described in \S \ref{scenario}
and \S \ref{fatima2}. The main results are
pointed out in \S \ref{results}, and their reliability is
discussed in  \S \ref{reliability}.
Finally, we analyze in \S \ref{conclusions} the consequences
to be drawn from the study.

\section{Observations\label{observations}}
\begin{figure}
\plotone{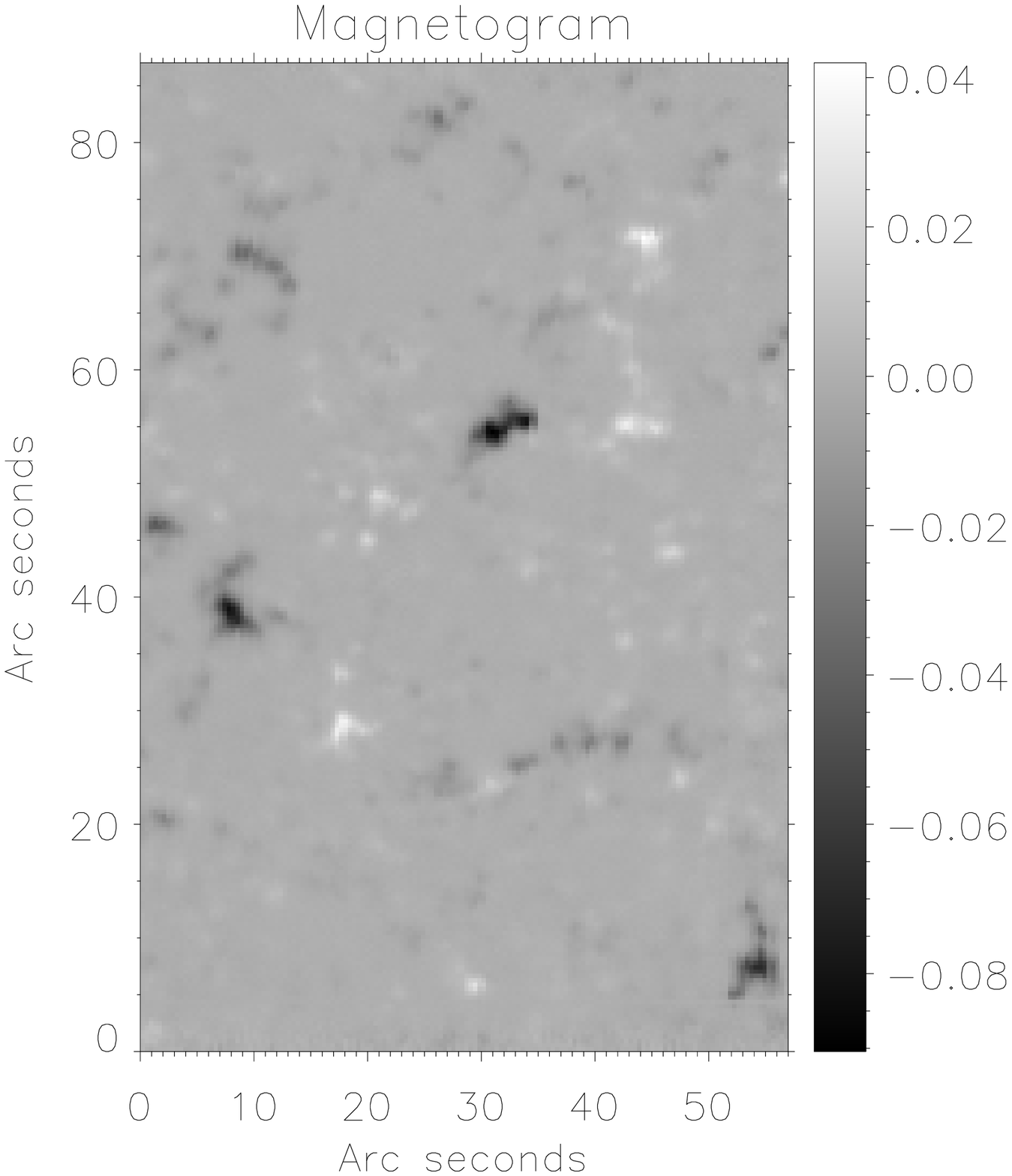}
\protect\caption[ ]{
Magnetogram of the observed region.
The network stands out clearly.
We display the degree of 
circular polarization in the blue wing of the Fe~{\sc i}
line at 6302.5~\AA . The vertical bar gives the equivalence between
grade of grey and polarization.
The spatial coordinates of the map are 
in arc seconds for the lower left corner.
}
\label{magnetogram}
\end{figure}

\begin{figure}
\plotone{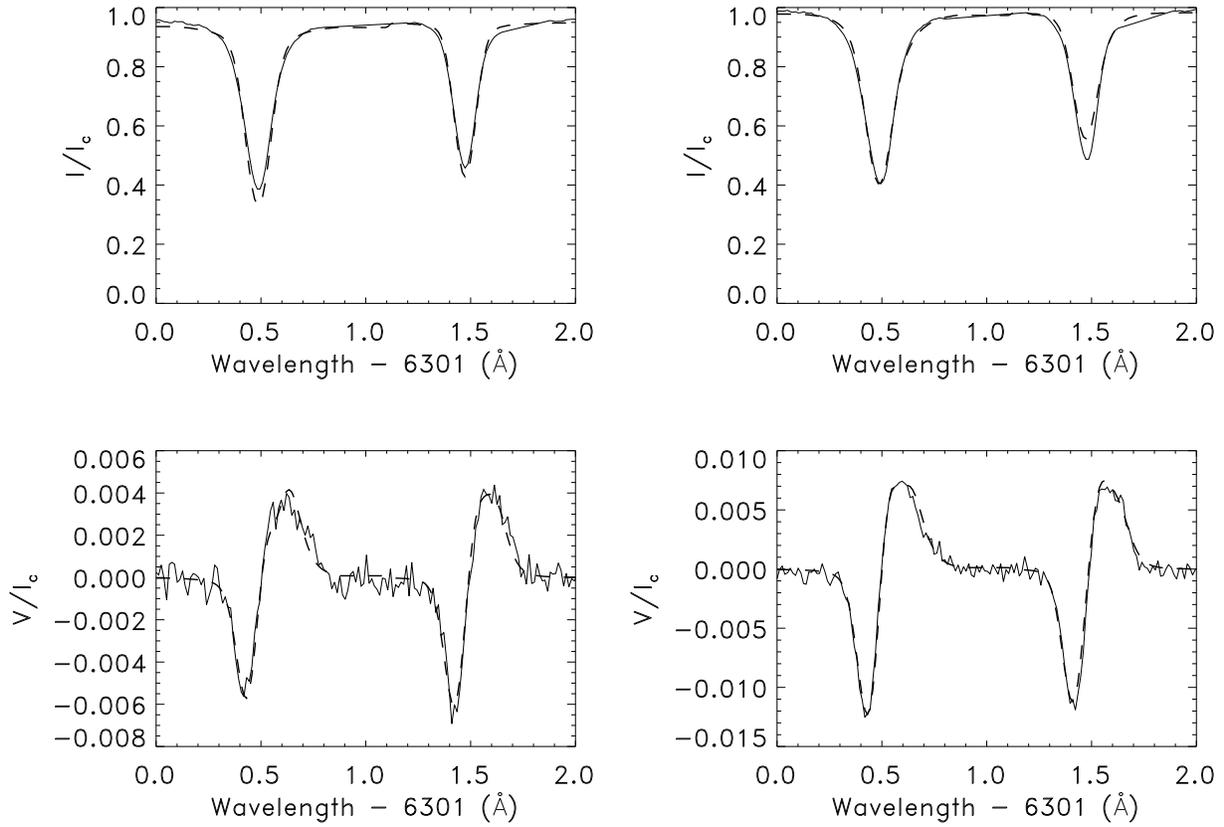}
\protect\caption[ ]{Examples of fits to profiles with relatively high
signals. 
Solid lines: observations. Dashed lines: fits.
Each column corresponds to a point on the Sun.
Stokes $I$ and $V$ are normalized to the quiet Sun
continuum intensity $I_c$.
Wavelengths are given in \AA\ from 6301 \AA.
}
\label{profs1}
\end{figure}

\begin{figure}
\plotone{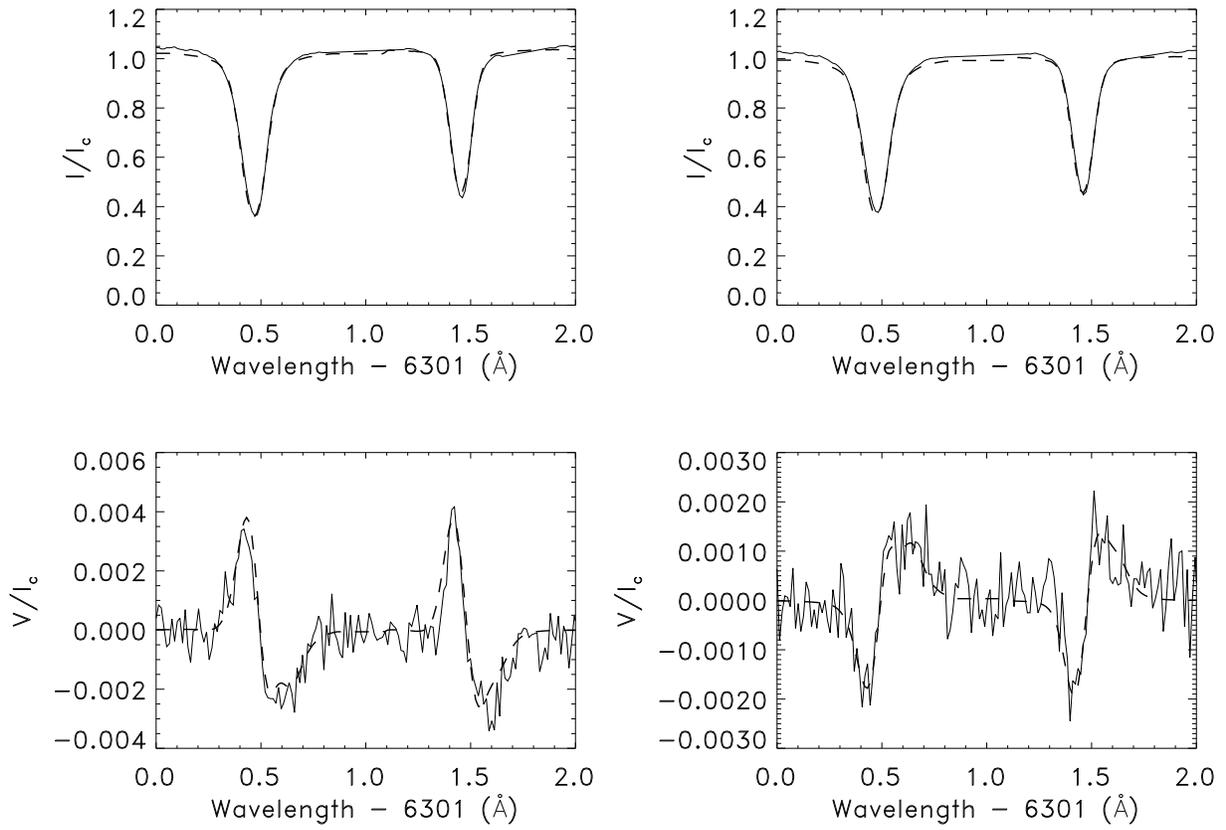}
\protect\caption[ ]{Examples of fits to profiles with very low
signals. Solid lines: observations. Dashed lines: fits.
For the other symbols, see Figure~\ref{profs1}.
}
\label{profs2}
\end{figure}

\begin{figure}
\plotone{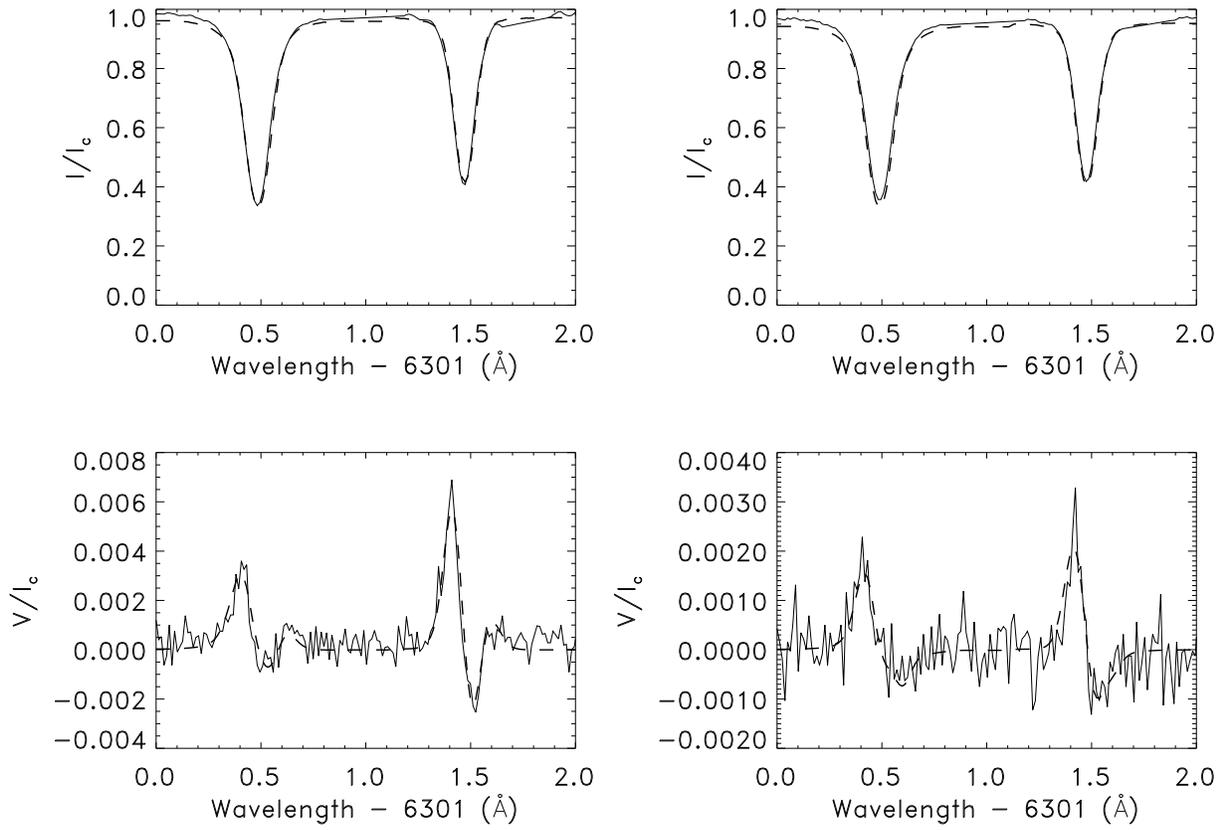}
\protect\caption[ ]{Examples of fits to extremely asymmetric profiles.
Solid line: observations. Dashed line: fits.
For the other symbols, see Figure~\ref{profs1}.
}
\label{profs3}
\end{figure}

	The observations are summarized for
the sake of comprehensiveness.
Further details can be found in 
\citeN{lit96} and SAL.
The polarized spectra were gathered with the
Advanced Stokes Polarimeter 
(\citeNP{elm92};
\citeNP{lit93}) working on
a very quiet 57\arcsec $\times$ 90\arcsec~
region near disk center (Fig.~\ref{magnetogram}).  Scanning the map 
with a 
sampling interval $\simeq$ 0\farcs 38
required some 9 minutes.
Stokes $I$, $Q$, $U$ and
$V$ profiles of the Fe~{\sc i} lines at 6301.5~\AA\ and
6302.5~\AA\
are available for every spatial position in the mapped region.
Various tests indicate that the map reaches an angular resolution
of about 1\arcsec~.
The noise of the Stokes spectra is better
than $10^{-3}~I_c$ , $I_c$ being
the continuum intensity (see Figs.~\ref{profs1},
\ref{profs2} and \ref{profs3}). The Stokes spectra
were corrected for instrumental polarization
to a level below the noise
(\citeNP{sku97}).
SAL 
inverted only those points of the region having a maximum 
Stokes $V$ signal
above an arbitrary threshold of
$3\times 10^{-3}~I_c$.
Some 5200 pixels or
15\% of the mapped area 
satisfy this criterion. 
In this paper we extend the inversion to the entire map, although our
analysis will be restricted to 
those points where the maximum Stokes $V$ signal is at least
twice the noise of each particular spectrum. Since the
noise of the different spectra varies,  this criterion 
optimizes the number of available spectra for a given signal to noise
ratio.
The noise is computed
in a continuum window where no polarization
is expected. We retain approximately 27\% of the map for analysis.  

\section{Inverting the Stokes profiles}

\subsection{The inversion code}
\label{fatima}

The Stokes data were inverted using
FATIMA (see \citeNP{soc01b}, hereafter SNLAL; \citeNP{ree00};
\citeNP{lop01}). This IC
is based on the Principal Component
Analysis of the profiles, which extracts the most relevant features
present in the spectra and compares them to a database of synthetic
profiles. The model atmosphere producing the synthetic profile in the database
that is closer to the observed one is chosen as the solution of the
inversion.

The main advantages and drawbacks of FATIMA with respect to other inversion
procedures are discussed in detail by SNLAL. In particular, there are three
important benefits that make it suitable for our purposes here:

\begin{enumerate}
\item Since it performs a global search over the whole space of possible
models and profiles, FATIMA never settles on secondary minima. In this sense,
it is very stable. If a discrepancy between the synthetic and observed
profile is found, it must be ascribed to either the adopted physical
scenario not being a valid description of the Sun or to the database
not sampling the model hyperspace in an adequate manner.
\item As shown in SNLAL, FATIMA is relatively less sensitive to noise 
than other inversion codes. This is also related to the stability issues
discussed above. Inversions based on iterative minimization algorithms are
more likely to find secondary minima when faced to profiles with very low
signal to noise ratio, like those analyzed in this work.
\item It is extremely fast, which results in the ability to process large
data sets in a very short time.
\end{enumerate}

An important point to bear in mind is that FATIMA does not solve the
radiative transfer problem, and is therefore not bound to any particular type
of model atmosphere. Instead, it simply relies on a precomputed set of models and
profiles that are stored in a database. This means that it may
be applied to any particular situation, independently of the physical
ingredients that need to be taken into account, provided that the database
has been constructed in a suitable manner. 

\subsection{The physical scenario \label{scenario}}

Although the emphasis in SNLAL was given on Milne-Eddington model
atmospheres, FATIMA 
may also be applied to more sophisticated scenarios. One only needs to
construct the database using profiles that are consistent with the model
assumed. To this aim we used the models obtained by SAL from the MISMA
inversion of the points with strongest polarization in the region
described in \S \ref{observations}.

The MISMA models consist of two magnetic components embedded in a
non-magnetic background. The two magnetic components are allowed to have
different magnetic field strengths, polarity, mass flows, 
occupation fractions (fraction of occupied atmospheric volume),
etc. These components are not monolithic structures. Instead, they
represent the mean properties of a conglomerate of many different
very narrow magnetic fluxtubes.
The temperature stratification is forced to be the same inside and outside
the fluxtubes. This is justified with the argument that
the radiative energy exchange
becomes an efficient smoothing mechanism that smears out any possible
thermal imbalance 
in seconds (e.g., \citeNP{spi57}; \citeNP{kne87}; \citeNP{sti91}).
Conservation of mass and magnetic flux is imposed in each magnetic
component. Hence the magnetic field becomes weaker as the tubes
spread with height, and the velocity amplitudes increase as the density
drops. The tubes and the surrounding environment are in horizontal mechanical
balance, i.e., at each height the magnetic pressure and gas pressure in the tubes
is
compensated by the gas pressure of the surroundings. Hydrostatic equilibrium
is imposed in order to determine the density stratification of the tubes and
the non-magnetic background. Since the velocities inferred from the
observations are sub-critical, this should be a fairly good approximation.
Finally, magnetic fields are assumed to be vertical, an assumption
that does not limit the validity of the models. Should the structure be
inclined, it produces the spectrum of a vertical structure except for
a
multiplying factor of the order of the cosine of the 
inclination (\citeNP{san99b}).

\subsection{Application of FATIMA\label{fatima2}}

The vertical magnetic  fields in our model atmospheres do not produce
linear polarization and therefore we do not consider Stokes $Q$ and
$U$ in the fits.

Some of the techniques given by SNLAL to reduce the dimensionality of the
problem are not applicable here because we are not dealing with
Milne-Eddington model atmospheres. In particular, the scaling of the source
function and the filling factor, which allows the IC to recognize similar
profiles with different amplitudes, is not applicable here. For this reason,
the database with the original synthetic profiles does not
represent a valid statistical description of the whole observation, since it
only contains the points with the strongest polarization signals. It 
exhibits an important bias, and FATIMA will be unable to provide
good fits to many profiles.
To overcome this difficulty, we have set a global magnetic filling factor as
a parameter that affects all MISMA components in the same way.
FATIMA selects the model atmosphere in the database
producing the synthetic Stokes profiles
$\{I_i,V_i\}$ that minimizes the residuals
\begin{equation}
\chi^2=\sum_i \Bigl\{\bigl[{{V_{oi}}\over{V^{m}_{o}}}- {{V_i}\over{V^{m}}}\bigr]^2 
+ \bigl[I_{oi}- I_i\bigr]^2\Bigr\}. 
\label{ff0}
\end{equation}
The index $i$ varies with wavelength,
whereas the set $\{I_{oi}, V_{oi}\}$ represents
the observations. 
The Stokes $V$ profiles, both synthetic and observed,
are normalized in equation (\ref{ff0}) to their extreme values
$V^m$ and $V_{o}^m$. 
The scaling factor is then given by
\begin{equation}
\alpha= V^m_o / V^m.
\label{ff}
\end{equation}
Once the best fitting profile is found in the database, the filling
factor of the individual MISMA components in the original model is
multiplied by the global filling factor (\ref{ff}) to render
the best fitting model atmosphere.

By accepting such scaling, we assume that the filling factors involved are
much smaller than unity, which is the case here. Moreover, it is also assumed
that the original spectra
inverted by SAL represent an adequate sample of all the line shapes
that may be
found in the entire map. They only exhibit stronger polarization because,
statistically, the magnetic components are present with a larger filling
factor in those points. 
This is a reasonable assumption since the classification 
into low or high signals carried out by SAL
does not obey physical differences, but it results from an arbitrary
thresholding relative  to the noise of the observations. 
In addition, 
the extrapolation is rather modest affecting polarization signals
that are at most a factor three weaker than the weakest in
the database. 
If this scaling does a proper job, then the modified
version of FATIMA should be able to provide good fits to the observations
over the whole map, including the line asymmetries. As we discuss
in \S \ref{quality}, this  is
indeed the case. All observations are successfully inverted with this
scenario, and we are able to obtain a model MISMA for each pixel.

The telluric lines in the
observed spectral region were removed from the intensity spectra before
proceeding with the inversion. This was
accomplished by interpolating between two points far in the wings of
these lines. Ideally, these distant points should be in the continuum. In our
case, however, the Fe~{\sc i} lines of interest are too close in wavelength,
which results in a slight deformation of the Fe line wings. This effect has
negligible impact on the inversions, because both the synthetic and
the observed profiles have been clipped in exactly the same manner, and also
because most of the information retrieved is contained in the Stokes~$V$
profiles, which are not affected by the telluric lines.

\section{Results}
\label{results}

When FATIMA runs on the observed map, we obtain a model MISMA
and a synthetic profile for each pixel. 
Then we select
only those points fulfilling the criterion described in \S \ref{observations}
which left approximately 27\% of the surface available for study.
For all these points we have the variation with height in the
	atmosphere of the relevant physical parameters (from
	magnetic fields to temperatures). Some  general
	trends of these properties are described in the 
	forthcoming sections.

\subsection{Quality of the fits\label{quality}}

The overall quality of the fits
is very good, in the sense that all observed profiles are well
reproduced within the noise level.  
Figures~\ref{profs1} to~\ref{profs3} show some
representative examples of different fits to observations with various signal
levels. In particular, the
synthetic profiles reproduce all sorts of Stokes $V$
asymmetries, a fact already known for the strongest polarization signals
analyzed by SAL, but which remains valid for the weak signals
studied here. 
Therefore, the assumption that our database is representative
of the whole data except for the effect of a global filling factor, is
justified {\it a posteriori} (see \S \ref{fatima2}). 
\

\subsection{Magnetic field strength\label{strength}}

Figure~\ref{field} shows a map of the average unsigned magnetic field strength
at the  base of the quiet Sun photosphere (i.e., the height
where the continuum optical depth at 5000 \AA\ 
equals unity or the total pressure becomes 1.3 $\times 10^5$ dyn cm$^{-2}$). We represent the mean 
field strength
of the two magnetic components in the resolution element
(a density-weighted mean that mimics 
the result to be obtained if a single field strength were assigned by the
IC; see \citeNP{san00c}).
The network may be easily recognized as patches
of strong field whose centers match the signals observed in 
the magnetogram
(Fig.~\ref{magnetogram}).  These patches are much larger
than their counterpart in the magnetogram, indicating 
a large decrease of occupation fraction towards the cell interiors
(see also Fig.~\ref{ffactor}).
Even outside the network most of the
pixels harbor magnetic fields stronger than 1 kG.
However, they tend to be slightly weaker than those around network patches.
Among the Inter-Network (IN) fields, we find a few intrinsically weak sub-kG fields.
The weakest fields tend to appear in the external parts of magnetic 
structures having larger field strengths, although the converse
is not true. Frequently magnetic structures become below the 
observational threshold with no transition through a weakening
of the field strength.

Figure~\ref{statistical} shows the magnetic field strength
versus the magnetic flux density of the observation.
The latter is defined as the total magnetic flux across the pixel
divided by the area of the pixel, and 
it is supposed to be
the quantity shown in magnetograms (e.g. 
	\citeNP{kel94}).
As one moves towards weak flux densities
(i.e., weak polarization signals), the mean
field slightly decreases and, simultaneously,
the scatter of possible values increases.
Still, most pixels show kG fields. 
We argue in \S \ref{reliability} that these field strengths 
are not an artifact
of the inversions but real. However, measurements with
the IR Fe~{\sc i} line at  15648~\AA\ convincingly show sub-kG
fields associated with the IN fields (see Fig.~\ref{statistical}),
which leads to the conclusion that both weak and strong fields
have to co-exist within regions 1\arcsec across. 
We consider this point in \S \ref{conclusions} below.

\subsection{Magnetic flux\label{mf}}

Figure~\ref{flux} shows the total (unsigned) magnetic flux of the region
when all those pixels above a given level of Stokes $V$ signal
are
considered. In agreement with SAL, the amount of magnetic
flux in the region keeps increasing as the sensitivity
improves, with no clear signs of saturation.
The total flux of the region in the pixels that we analyze amounts to
4.8 $\times 10^{20}$ Mx.  
The mean flux density, corresponding to the total flux over the
total area of the region, is 18 G. 
If one employs only points outside the network (defined as those with a
occupation fraction smaller than 10\%), the flux density
decreases to 10 G. 

These values for the magnetic flux and magnetic flux density are
large in various senses.
SAL obtained for the same region
3.3 $\times 10^{20}$ Mx when
only the strongest signals are considered. We gain a factor two
in sensitivity which produces a 50 \% increase of the magnetic flux.
It is clear how the sensitivity of the observation
plays a crucial role
in the determination of the quiet Sun magnetic flux budget (and
therefore the energy budget).

On the other hand, we also computed the flux density of the
map from the
magnetogram in Figure~\ref{magnetogram} following the standard 
procedure to calibrate magnetograms\footnote{We use equation (\ref{wfa})
below with the derivative computed using
the mean Stokes $I$ profile observed in the entire region. This leads to 
a calibration constant of 5145 G per Stokes $V$ signal unit.}.
The total flux thus obtained is 2.0 $\times 10^{20} $ Mx, which infra-estimates the flux
by a factor 2.4. This deficit can be pinned down to various
factors:
the saturation of the Fe~{\sc i}~6302.5~\AA \, line whose circular
polarization is no longer proportional to
the magnetic flux for kG fields (e.g. \citeNP{ste73}), 
the weakening of the
lines in magnetic concentrations (\citeNP{har69}), but also to the
partial 
cancellation of the existing flux. Two opposite polarities frequently
coexist in the  
resolution elements. Figure~\ref{mixed}
shows the fraction of pixels where two opposite
polarities in the resolution elements are invoked to reproduce
the observed polarization.

The amount of flux in the quiet Sun
is also large as compared to the flux observed in active regions
during the solar maximum.
Should the full Sun be
covered by quiet regions like the one  we study, the
total unsigned flux across the solar surface would be
$10^{24}$ Mx,
which is similar but larger that the magnetic flux during the
solar maximum measured with standard low spatial
resolution magnetographs 
(see \citeNP{sch94}, Fig.~3). This fact indicates that 
taking into account the contribution of the IN fields
may be important to evaluate and characterize
the flux budget during the cycle, including the evolution with time.

\subsection{Mass motions\label{veloc}}

The observed asymmetric
Stokes $V$ profiles (Figs.~\ref{profs1}, \ref{profs2} and
\ref{profs3}) follow from a rather simple velocity field.
The magnetic components are
embedded in a downflowing non-magnetic environment.
The two magnetic components exhibit different motions.
One of them is almost at rest and contains most of the
magnetized mass. The other one tends to undergo strong downflows.

The distribution of velocities at the height corresponding
to the base of the
non-magnetic photosphere is
shown in Figure~\ref{vel}. It contains the mean and the standard
deviation as a function of the flux density.
We separate the two components according to the velocities.
The mean of the distribution in the slow component (the solid line)
is very small for large
magnetic flux densities. In fact, it is smaller than the
uncertainty given by the absolute wavelength calibration (refer
to SAL).
The scatter is also small in this region of
the diagram. 
As we move to
lower flux densities 
(say below 10~G), the distribution shifts to negative (upwards-directed)
velocities and the scatter increases slightly.
The second component (the dashed line) is
characterized by important downflows of $\simeq$~3~km~s$^{-1}$ at this height. Its
mean value remains approximately constant over the whole range of flux
density values, but in this case the scatter grows significantly when we
consider weak signals.

\subsection{Occupation fraction\label{of}}

Figure~\ref{ffactor} shows the occupation fraction, i.e., the 
fraction of volume in the resolution element  occupied by magnetic
fields. The map corresponds to the lower photosphere. Occupation
fractions span from some 20\%, for
a few network concentrations, to 0.5\%, for the weakest polarizations
analyzed here. The mean value for the surface that we analyze is about 4\%. 
If the network concentrations are not considered 
(a discrimination based on the polarization threshold used in \S
\ref{mf}), it decreases to 3\%. If one considers the whole region
instead of the 27\% having reliable signals, then the magnetic fields
that we detect occupy only 1\% of the total volume. This figure is 
a lower limit because it assumes that places with negligible
signal contain no magnetic field.

%\old{
%Could we give a figure for the fraction of photospheric
%volume occupied by the magnetic fields that we detect?
%This will satisfy Javier and others. Obviously it would be
%an lower limit. Point to mention Fig.~\ref{ffactor}.
%
%* Filling factor promedio sobre todo el mapa (considerando ff=0 para los
%puntos "no fiables"): 1\%
%
%* Filling factor promedio cuando sólo promediamos los puntos "fiables"
%(los demás no se consideran en el promedio): 4\%
%
%* Filling factor promedio fuera de la network cuando sólo consideramos
%los puntos "fiables": 3\%
%}
%
 
\section{Reliability of the results\label{reliability}}
This work employs an unusual method of inversion 
to analyze low polarization signals. 
Since the reliability of our conclusions is 
very much based on the dependability of the inversions,
we feel compelled to explain in some detail the
studies that give us confidence on the results.

First, we compared the model atmospheres
chosen by FATIMA for those points used to produce the
database. In 90\% of these points, the field strength retrieved by
FATIMA differs less than 10\% \, from the value in the database, which
proves that the IC is working properly 
in the five thousand cases where we can test it directly. 
In
particular, it shows that the principal component
decomposition of the profiles does not seem to introduce
artifacts.

Second, we wondered whether the noise of the observation was too
high so that the IC has no information and
it assigns model atmospheres
at random. Several arguments discard such concern.
The probability of finding a given
physical property in the whole region is not the
same as that obtained from the points
in the database. Should atmospheres in the database were
chosen at random, the distributions of magnetic field strength, velocities,
etc, would have to  mimic the original one. However, this is not the case.
Magnetic fields are sensibly weaker
in the whole region, the fraction of
opposite polarities larger, and so forth.
Figure~\ref{mixed} shows the fraction of times that
FATIMA assigns  mixed polarities in the
resolution element. It becomes 35\% in the weakest
signals analyzed here whereas, in average, only 13\% of the times
such mixed polarities are present in the original database.

Another hint that the IC is not assigning magnetic properties
at random can be inferred from Figure~\ref{field}. Note that neighboring
pixels show similar field strengths although the inversions of all of them
are independent. Since adjacent pixels show similar profiles,
this indicates the IC has been able to distinguish their
shapes and separate them.

Third, one of the important conclusions of the work is the
ubiquity of kG field strengths even outside the network. 
Since most of the
magnetic fields in the database correspond to kG
(SAL, Fig.~14), one
naturally wonders whether this bias determines the conclusion.
It does not. The database also contains sub-kG 
fields, but most of the 
times they are left aside on purpose by the IC.
There is enough information in the observed
profiles to distinguish between kG and sub-kG fields.
Most of the profiles observed in IN regions
cannot be produced by sub-kG fields. An example of how
the fits worsen if one attempts sub-kG field fittings in a
region whose best fit requires kG is shown by \citeN{san01c}, \S 4.
Here we offer a much simpler argument pointing in the same
direction. Suppose that the IN fields were covered by intrinsically 
weak fields of the kind inferred in the IR (\citeNP{lin99};
\citeNP{col01}). These fields are weak in two different senses. 
The radiative transfer can be
treated in the so-called {\it weak field approximation}, i.e., 
the Stokes $V$ scale with the wavelength derivative of
the intensity profile, 
\begin{equation}
V(\lambda)=-k g_{eff}\lambda_0^2 B {{dI(\lambda)}\over{d\lambda}},
	\label{wfa}
\end{equation}
where $k$, $g_{eff}$, $\lambda_0$ and $B$ have their usual
meanings (a constant, the effective
Land\'e factor, the mean wavelength of the spectral line,
and the longitudinal component of the magnetic field, respectively).
On the other hand, the magnetic fields are {\it dynamically weak} so that
they cannot determine the thermodynamic properties 
of the atmosphere. This second argument points out that in
weak magnetic structures, the thermodynamic properties
have to be close to those of the quiet Sun, so that
the relative strengths of Fe~{\sc i}~6302.5~\AA\ and
Fe~{\sc i}~6301.5~\AA\ are those
observed in the quiet Sun. Consequently,
for intrinsically weak fields, the Stokes $V$ signals have
to be given by equation (\ref{wfa}), $I$ being the quiet Sun spectrum $I_q$.
Figure~\ref{weak8} shows the product  $g_{eff} dI_q(\lambda)/d\lambda$.
It is clear how the signal to be expected in Fe~{\sc i}~6301.5~\AA\ is 
much weaker than that in Fe~{\sc i}~6302.5~\AA, as it is indeed observed
in some places (see Fig.~\ref{profs3}). However, observations usually  show
the two lines with similar degrees of polarization (Fig.~\ref{profs2}),
proving that the lines are formed outside the
weak field regime, which corresponds to kG. 
Figure~\ref{weak9} displays the scatter
plot of the maximum Stokes $V$ signals observed
in the two Fe~{\sc i} lines employed in the study.
Note how most of them appear clearly above the
line corresponding to weak fields, as deduced from the
derivatives in Figure~\ref{weak8}.
In short, despite the uncertainties of the inversion,
the presence of kG fields in the IN is a solid result. 

\section{Conclusions}
\label{conclusions}

We investigate the properties of quiet Sun regions
having very small magnetic flux (down to the smaller that we can
detect, equivalent to a few times 10$^{16}$ Mx;
see Fig.~\ref{statistical} and Fig.~\ref{vel}).
These regions, frequently referred to as ``non-magnetic'' regions,
occupy most of the solar 
surface  and therefore they may easily  contain a sizeable fraction of the
photospheric magnetic flux and energy (e.g., \citeNP{ste77};
	\citeNP{yi93}; \citeNP{san98c}).
Both observations and theory suggest
that the fields of the quiet Sun have 
a very complex topology (e.g. \citeNP{cat99a}; SAL), which
hinders an easy detection.
The tangling of field lines 
cancels to a large
degree
the polarization that
would reveal the presence of magnetic 
structures\footnote{Either as a true cancellation
of the Zeeman-induced polarization, or due to the intrinsic
weakness of the scattering polarization signals 
used to infer magnetic fields via the Hanle
effect.}. 
Many of the weakest signals appear in the network
cell interiors; the so-called Intra-Network fields (IN).
Although the first detections of weak IN fields
date back to the seventies (\citeNP{liv71}; \citeNP{liv75};
	\citeNP{smi75}),
we still have a very rudimentary knowledge of their basic
properties. Low angular resolution magnetograms have provided
information on 
their apparent shapes, magnetic fluxes and   
macroscopic motions 
(e.g., \citeNP{wan95}; \citeNP{zha98}, and references therein.)
However,
the advent of a generation of
spectro-polarimeters with improved sensitivity
has provided new data that remain to be incorporated into
a consistent paradigm.

	One of the first and important results of the
new polarimeters
has been finding very asymmetric Stokes profiles
(\citeNP{san96}; \citeNP{gro96}; \citeNP{sig99}; SAL).
Obviously, the observed polarization is produced in atmospheres 
that require
more than a single magnetic field and a single velocity 
to be described. The inference of the physical properties
of such atmospheres requires special inversion techniques.
This inference presents an additional complication,
namely, the observed polarization
signals are low and therefore close to the noise level.
In this work we propose, essay  and then use a new technique that copes
with these two difficulties. It reproduces  the asymmetries
of all observed Stokes profiles (\S \ref{quality}).
The application of the 
technique allows us to obtain model atmospheres for 27\% of the 
observed quiet Sun region.
The procedure is a modification of FATIMA (SNLAL; see \S \ref{fatima}
and \ref{fatima2}) which employs
a database of synthetic observations based on the
MISMA inversion of the strongest signals in the
region (SAL). This inversion technique is extremely
fast, and it offers a real chance to handle large data sets
of very weak asymmetric polarization signals,
like those presumably needed for the analysis of 
future observations.

\begin{figure}
\plotone{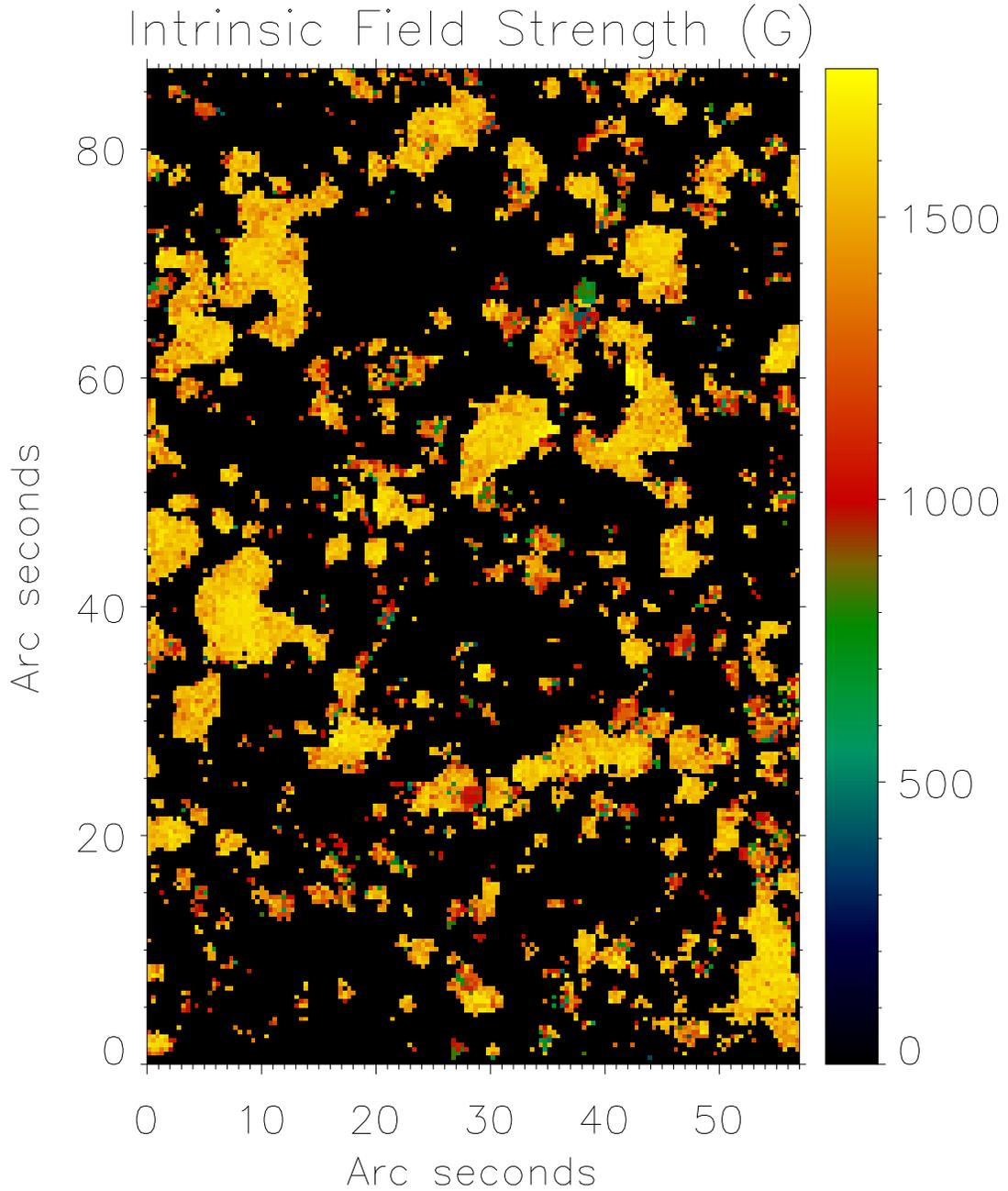}
%\plotone{/scratch2/texto/papers/paper34/f5d.eps}
%\plotone{/scratch2/texto/papers/paper34/f5e_psd.eps}
	%\includegraphics[width=\textwidth]{f5.eps} % referee
	%\includegraphics[width=0.45\textwidth]{f5.eps} % 2-column
	%\includegraphics[width=.55\textwidth]{field.eps} % referee
\protect\caption[ ]{Average unsigned magnetic field strength
at a height corresponding to the base of the unmagnetized
photosphere (i.e., where 
the total pressure amounts to 
1.3 $\times 10^5$ dyn cm$^{-2}$).
Even inside the network cells, most of the pixels harbor kG field strengths.
The vertical bar shows the equivalence between colors
and field strengths. 
Points with polarization below the threshold have been set to
zero.
The spatial coordinates of the map are 
in arc seconds.
}
\label{field}
\end{figure}

\begin{figure}
\plotone{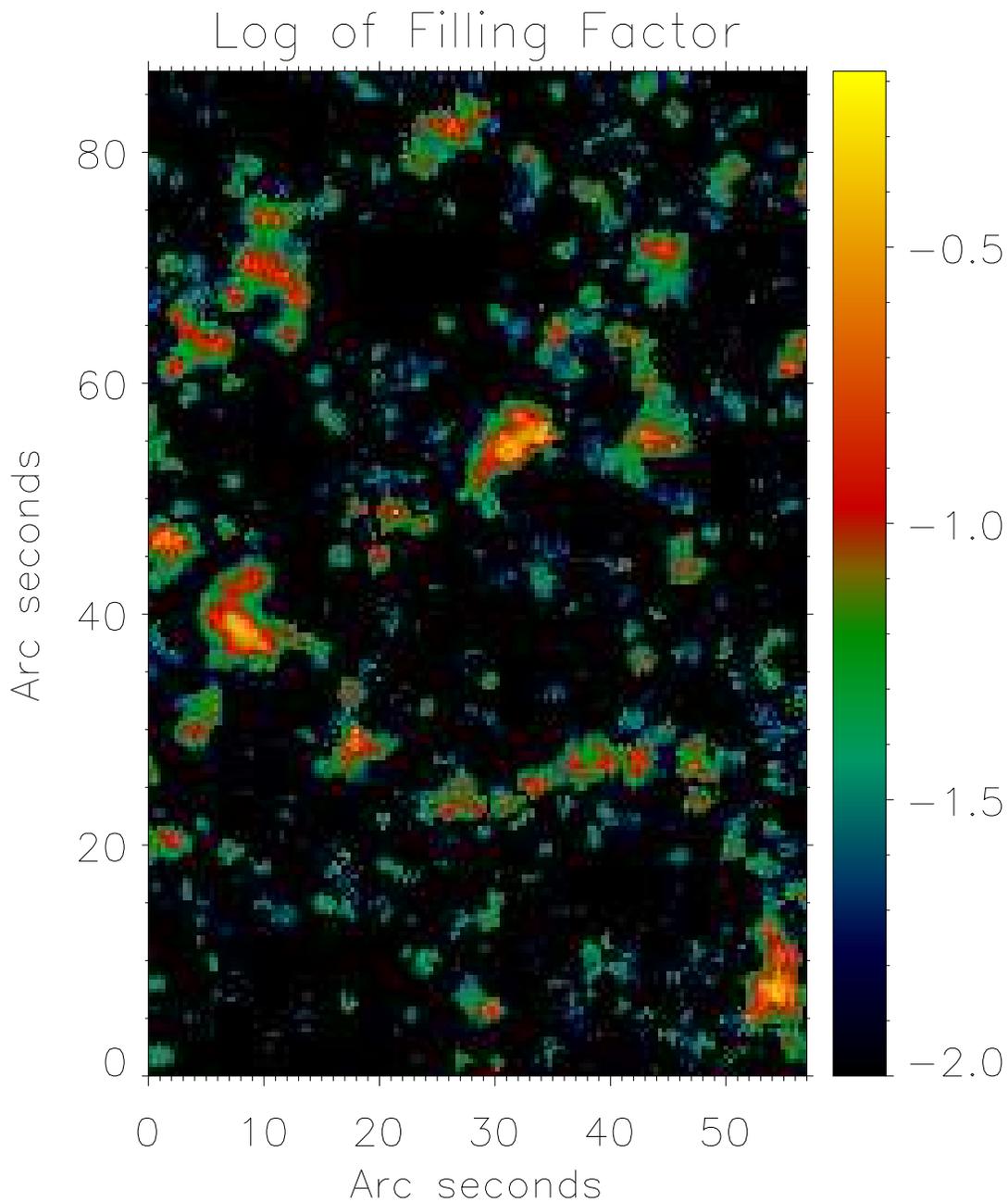}
%\plotone{/scratch2/texto/papers/paper34/f6d.eps}
%\plotone{/scratch2/texto/papers/paper34/f6e_psd.eps}
	%\includegraphics[width=\textwidth]{f6.eps} % referee
	%\includegraphics[width=0.45\textwidth]{f6.eps} % 2-column
	%\includegraphics[width=0.55\textwidth]{ff.eps} % 2-column
\protect\caption[ ]{Fraction of atmospheric volume occupied
by magnetic fields
at a height corresponding to the base of the unmagnetized
photosphere 
(i.e., where 
the total pressure amounts to 
1.3 $\times 10^5$ dyn cm$^{-2}$).
The vertical bar shows the equivalence between colors 
and occupation fractions, which is  given in a logarithmic 
scale
to evidence IN fields.
Points with polarization below the threshold have been set to -2.
The spatial coordinates of the map are 
in arc seconds.
}
\label{ffactor}
\end{figure}

\begin{figure}
\plotone{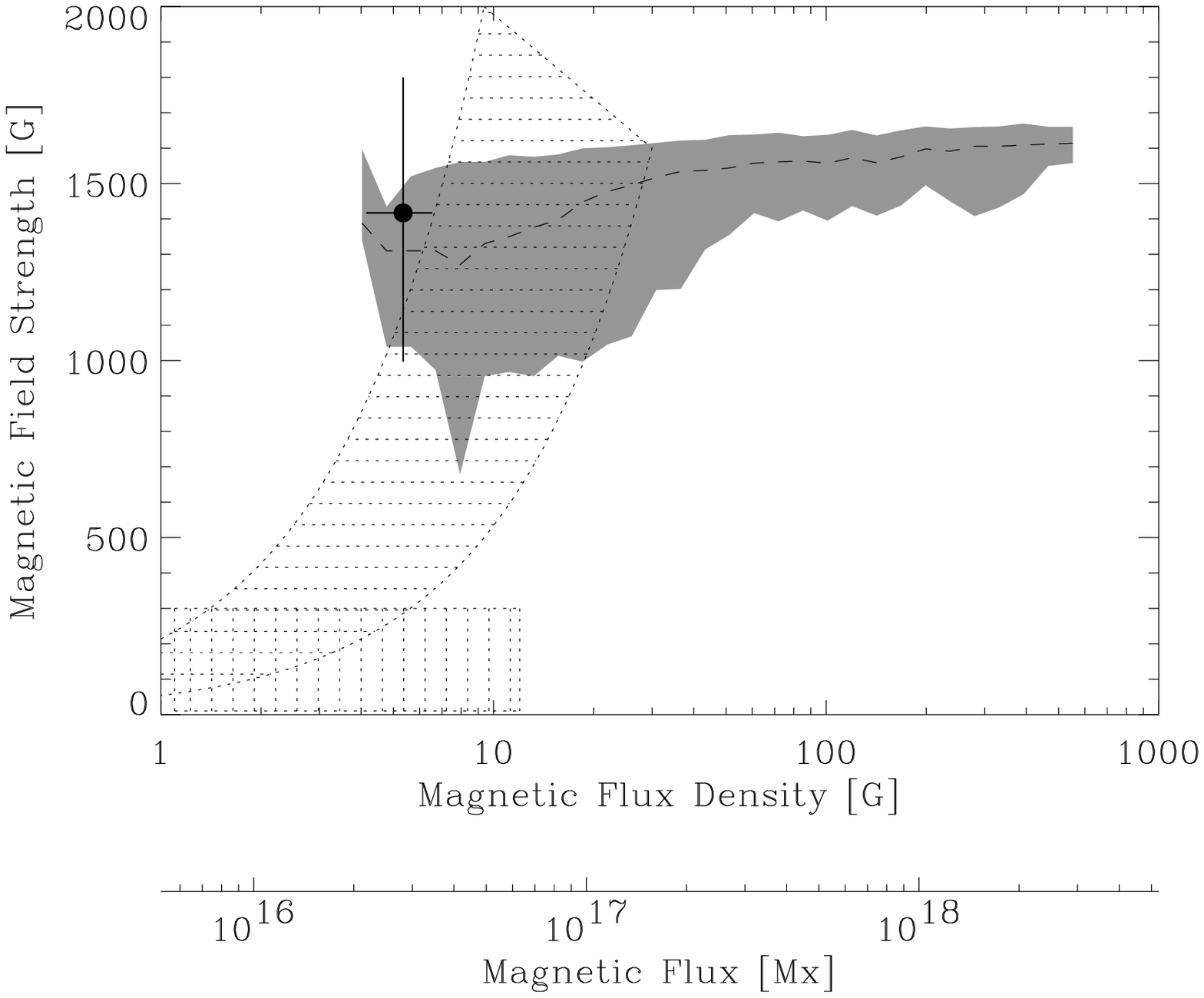}
\protect\caption[ ]{
Magnetic field strength as a function of
the magnetic flux density.
The horizontally hashed region corresponds to the limits observed
using the Fe~{\sc i} IR line at 15648~\AA\ by \citeN{lin95} and
\citeN{lin99}. 
The box in the lower left corner marks the results
independently obtained by \citeN{col01} using
the same line 
(the region in the plot includes 80 \% of the observed signals).
The analysis of this work leads to the shaded region,
which represents the mean (dashed line) as well as the standard
deviation above and below the mean (shaded area).
The point with error bars
was derived by \citeN{san01c}, using part of the dataset analyzed here.
The scale of magnetic flux has been computed assuming an
angular resolution of 1\arcsec.
}
\label{statistical}
\end{figure}

	The inversions show that most polarization signals
in the quiet Sun
correspond to kG fields at the base of the photosphere (\S
\ref{strength} and Fig.~\ref{field}). This is in agreement 
with previous observations based on Fe lines in the visible spectral range
(\citeNP{gro96} ; see, however, 
\citeNP{kel94}), and we show this conclusion to
be a solid result of our analysis (\S \ref{reliability}).
On the other hand, different observations based on the 
Fe~{\sc i} IR line at 15648~\AA\ almost exclusively
find sub-kG fields in regions having similarly low flux densities
(\citeNP{lin95}; \citeNP{lin99}; \citeNP{col01}).
Again this is a robust conclusion mostly based
on the separation of the Stokes $V$ peaks whose
small splitting
does not correspond to kG fields.
We believe that the apparent contradiction between visible
and IR based field strengths is not real.
As explained by SAL (\S 4.5),
the IR observations are not sensitive to the kG
fields that may exist. 
The two seemingly 
inconsistent  observations actually indicate that
various strengths often  coexist
within the resolution elements. 
According to the same observations,
kG and sub-kG fields fill only a small fraction of the photospheric
volume. In our case the detected magnetic fields occupy only 1\% of the
total volume (or 4\% of the volume if we restrict the estimate
to those places with detected signals; 
see \ref{of}). In IR observations the fraction increases since
the intrinsic field strengths are lower,
but still there is enough space for extra weak fields. In particular,
those inferred in Hanle effect based  determinations
of turbulent fields (e.g, \citeNP{ste98}; \citeNP{bia98}; \citeNP{bia99}).

\begin{figure}
\plotone{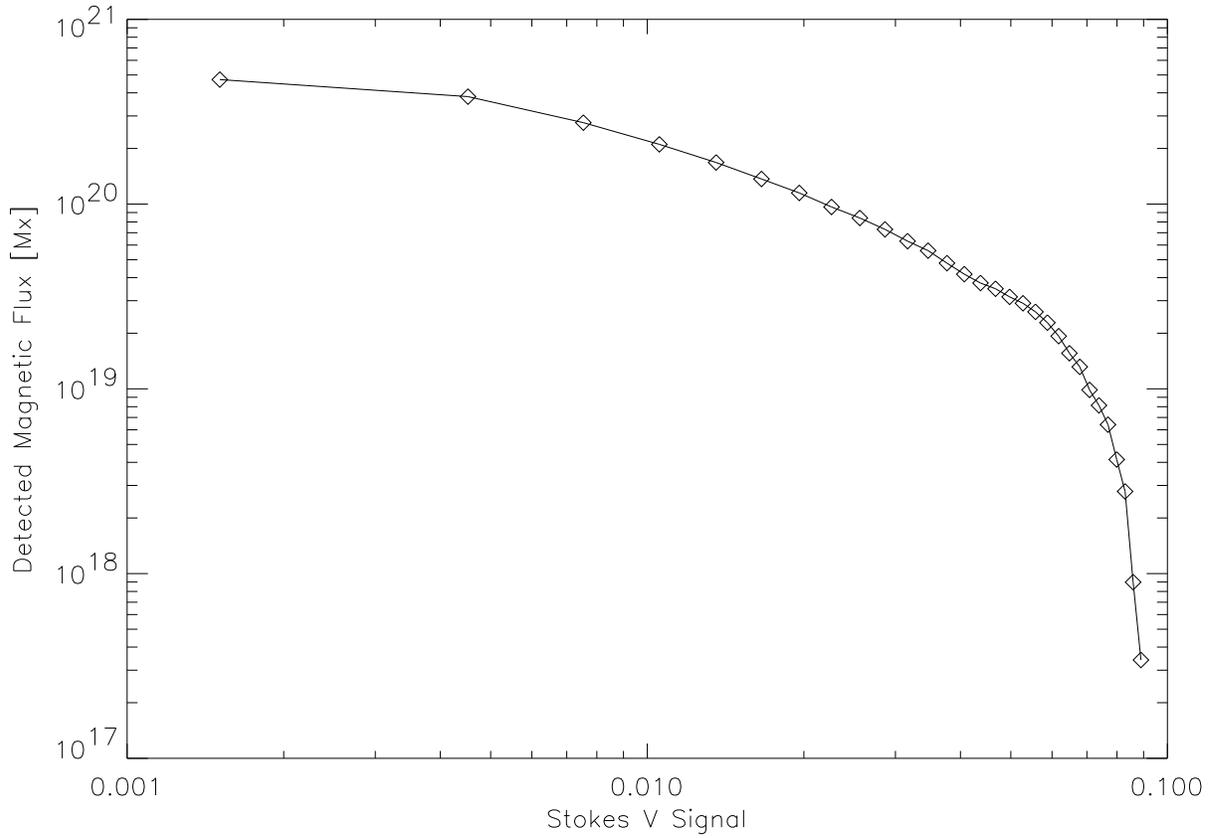}
\protect\caption[ ]{Magnetic flux in the region
	considering all the
	pixels with signals stronger that the abscissa.
	The total unsigned flux steadily increases as one considers
	weaker signals,
	with no clear signs of saturation for the weakest ones
	analyzed here. The Stokes $V$ signals are measured
	in the same units as the magnetogram in Figure~\ref{magnetogram}.
	}
\label{flux}
\end{figure}

\begin{figure}
\plotone{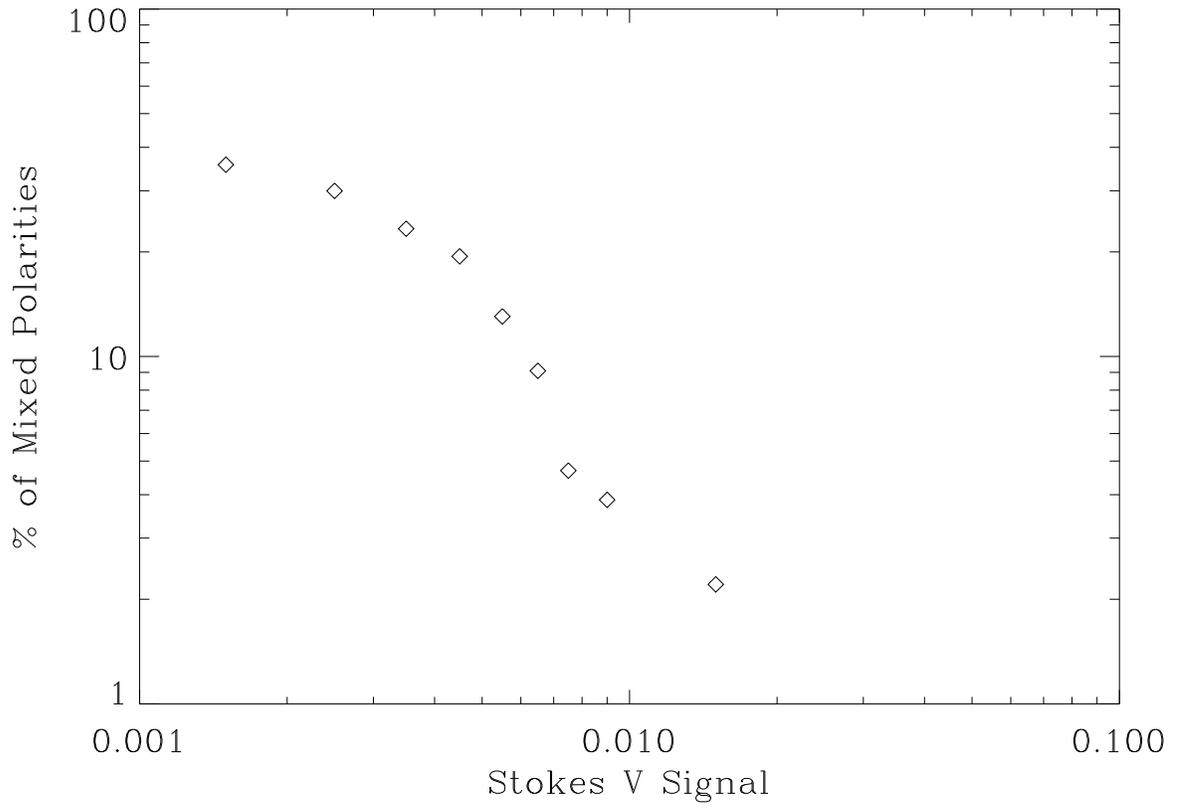}
%\plotone{mixed.eps}
\caption{
	Percentage of pixels having two different polarities in the
	resolution element as
	a function of the Stokes $V$ signal. The percentage
	reaches some 35\% for the weakest polarization
	signals analyzed here.}
\label{mixed}
\end{figure}

\begin{figure}
\plotone{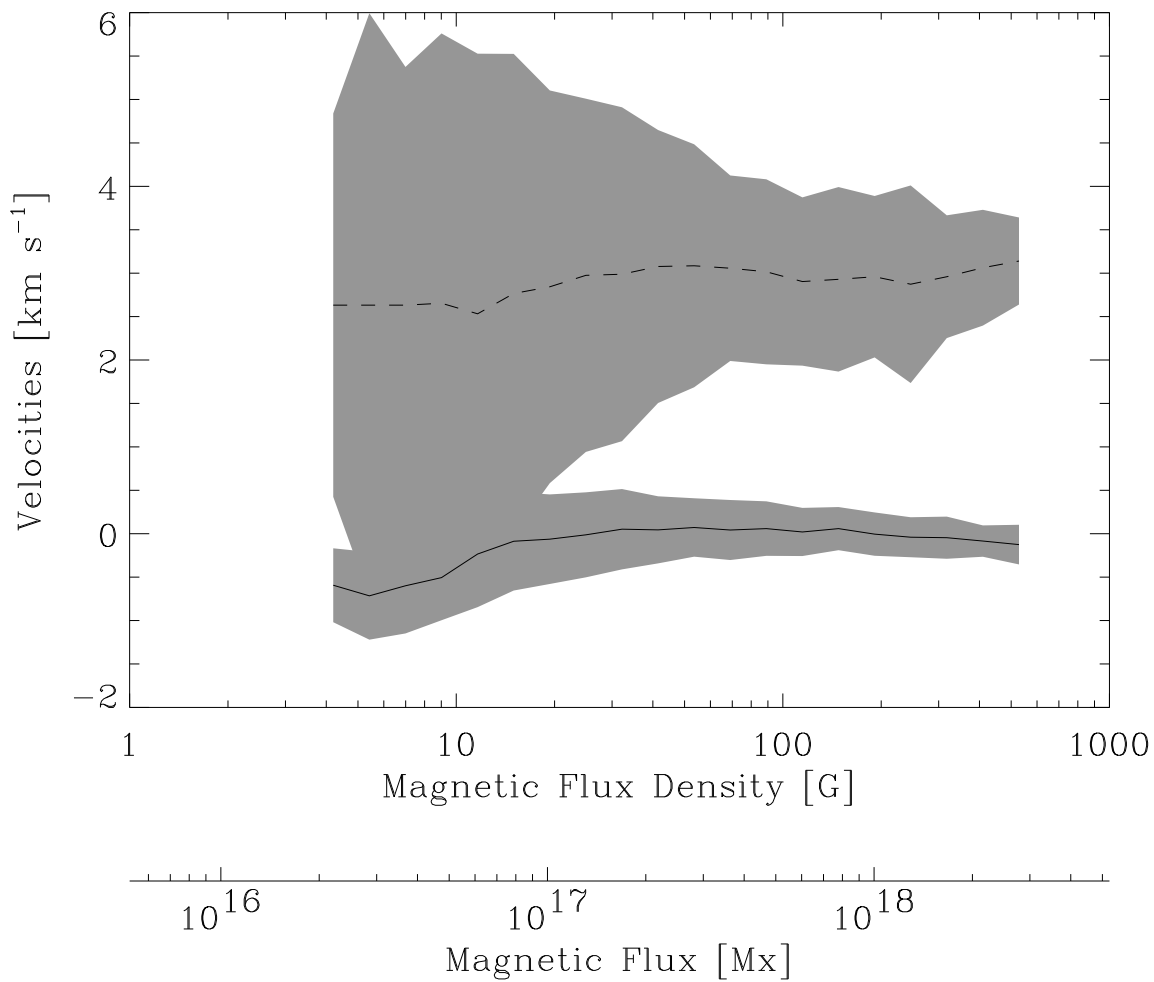}
%\plotone{pinta4.eps}
\caption{
The solid (dashed) line represents the mean of the velocities in the first
(second) magnetic component at a height corresponding to the
base of 
the quiet photosphere. It is represented versus the unsigned magnetic flux
density in the resolution element.
The shaded areas correspond to the mean plus-minus the standard deviation of the distribution.
Positive velocities are red shifts or downflows.
The magnetic flux axis has been computed as for Figure~\ref{statistical}
}
\label{vel}
\end{figure}

\begin{figure}
%\plotone{weak8.eps}
\plotone{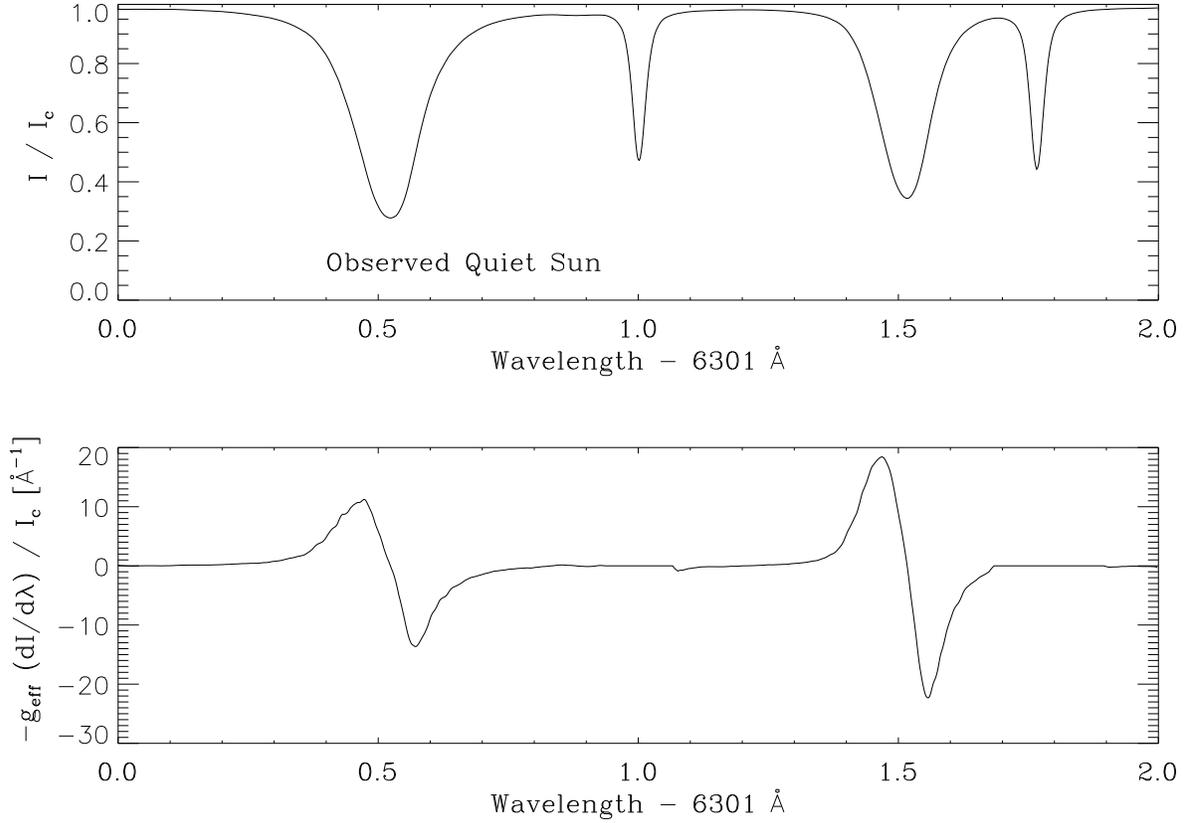}
\caption[ ]{Top: Stokes $I$ profiles of Fe~{\sc i}~6302.5~\AA\ and 
	Fe~{\sc i}~6301.5~\AA\
	observed in the quiet Sun at the solar
	disk center (\citeNP{del73}). If the spectral lines 
	are formed under weak magnetic field conditions,
	the relative amplitude of the Stokes $V$ signals should
	be given by the weak field approximation, i.e., 
	the derivative of the Stokes $I$ profiles times
	the effective Lande factor of the line (bottom).
	The signal to be expected in Fe~{\sc i}~6302.5~\AA\
	is some 60\% larger than that produced by Fe~{\sc i}~6301.5~\AA .
	Wavelengths are given in \AA\ from 6301~\AA.} 
	\label{weak8}
\end{figure}

\begin{figure}
%\plotone{weak9.eps}
\plotone{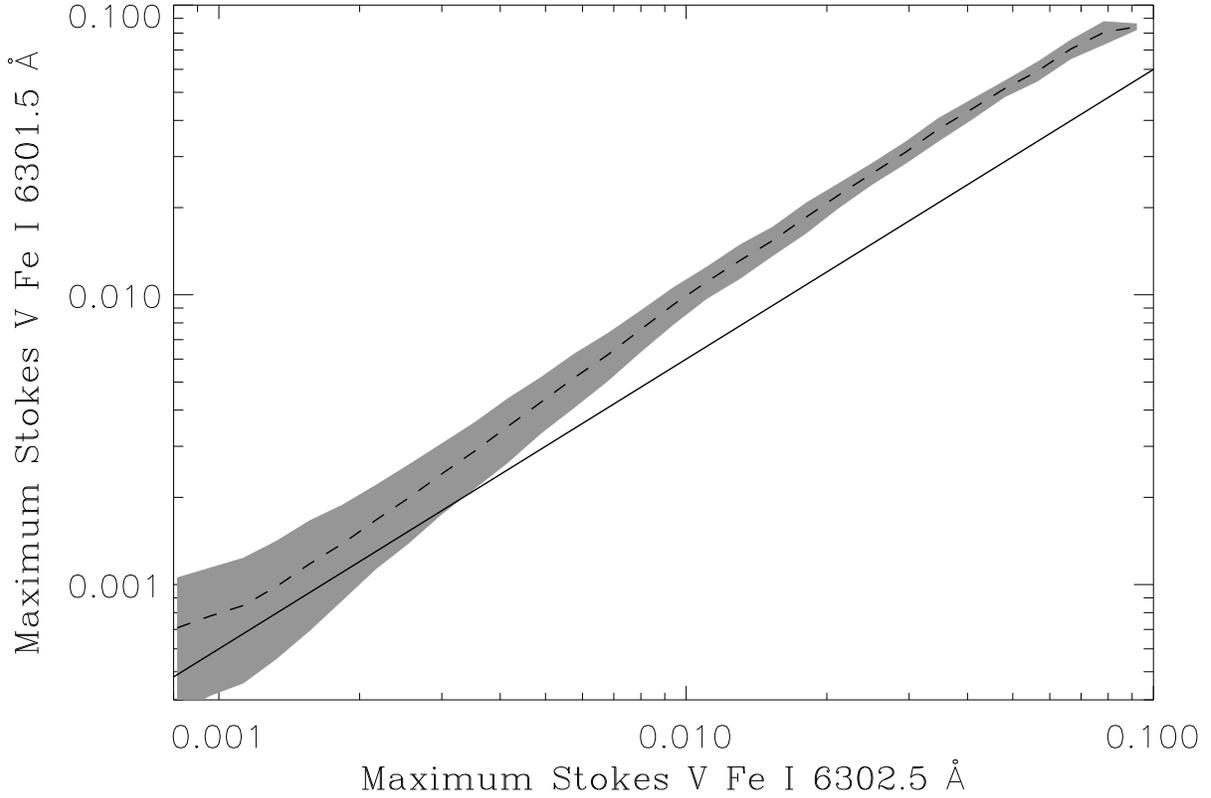}
\caption[ ]{Scatter plot of the maximum Stokes $V$ signals observed in 
the two Fe {\sc i} lines subject to analysis. The shaded region
corresponds to the mean plus and minus the standard deviation among the
Fe~{\sc i}~6301.5~\AA\ signals with a given
Fe~{\sc i}~6302.5~\AA\ signal. The mean of this distribution
is represented by the dashed line.
The solid line shows 
the relationship to be expected
if the magnetic field were intrinsically weak (see Fig.~\ref{weak8}). Most of the
observed ratios correspond to strong fields, i.e., to points above this
ratio.}
\label{weak9}
\end{figure}

The scenario described above is very much in favor of a magnetic field
with complicated topology, whose properties we are just
beginning to glimpse. 
Two additional 
results of our study have to be included
as part of the complexity.
First, we find that a large fraction of the signals 
correspond to
two different polarities in the resolution element
(35\% of the weakest signals, see Fig.~\ref{mixed}). 
The magnetic field lines of the two magnetic components
in the model atmosphere cross the horizontal plane in
opposite senses.
Second, the bulk magnetic flux of the region
is contained in the weakest signals that we detect (see Fig.~\ref{flux}).
Moreover,
the flux that we find is larger than the fluxes
associated to the IN fields in previous works
having lower sensitivity (\S \ref{mf}),
indicating that the amount of flux that we detect still is a
matter of sensitivity.
In other words, a fraction of the existing magnetic structures 
is missing. They
will emerge in future measurements  with 
improved 
sensitivity and better spatial resolution analyzed through the
proper 
diagnostic technique.

Should the quiet region that we observe be typical of the
rest of the Sun, its flux would be
comparable to (and even larger than) the flux
crossing the solar surface in the form of active regions
at any time
during the solar cycle maximum (some $10^{24}$ Mx). 
Do quiet Sun fields  play an active role during the cycle?
We simply do not know. However, 
since the flux (and presumably energy) that they contain
is so important, it is difficult to discard any potential influence.
Keep in mind that we detect 
only a fraction of the existing flux.

This work uses
a new
inversion technique
to extend
the study by SAL
to weaker signals.
The conclusions
expressed here are very much in the line of those already
found by SAL, except that here we rest on a larger statistical
basis. In particular, the velocity flows that we find (\S \ref{veloc})
are also of the
kind that fits the typical plage and network
asymmetries in a MISMA context (\citeNP{san97b}). Most of the 
magnetized mass is at rest but there is also
strong co-spatial downflows that produces 
the observed asymmetric line shapes.

	The mechanism that concentrates magnetic fields to form
	kG fields remains unclear.  
We do not detect weak field strength precursor regions with enough magnetic flux to produce 
typical network magnetic concentrations ($10^{17}$ Mx -- $10^{18}$ Mx).
\citeN{san01c} proposed that the concentration
process takes place in patches of much smaller fluxes
($10^{16}$ Mx),
which are then gathered by the super-granular converging
flows to create the regular network structures that
we observe. If so, there should be a population of structures
of low flux but already concentrated kG fields. 
We believe that they begin to show up in the upper left
corner of Figure~\ref{statistical}. The finding of such population demands 
the existence of concentration mechanisms able to efficiently 
operate with low flux structures,
like the thermal relaxation process described by
\citeN{san01c}. 

As we pointed out above,
the investigation of the quiet Sun fields is still in a developing phase.
Each new contribution suggests observations
to either confirm previous findings or explore conjectures.
In this sense, the above results naturally convey
to simultaneous and co-spatial IR and visible observations.
They will
prove or discard the co-existence of sub-kG and kG field
strengths that we support. 
They also would allow
to witness the concentration process of low flux structures.
Such observation requires high sensitivity
to both kG and sub-kG field strengths, 
which can be attained by the
simultaneous use of IR and visible spectral lines.
On the other hand, we conjecture that the quiet Sun fields may participate
in the global solar magnetic cycle playing a significant
but still unknown role. Monitoring of the  quiet Sun magnetism 
along the cycle using sensitive
spectro-polarimeters is eagerly needed\footnote{
This may be an ideal task for the vector magnetograph of the
synoptic instrument SOLIS;
	\url{http://www.nso.noao.edu/solis/}.
	%\anchor{http://www.nso.noao.edu/solis/}{SOLIS homepage}.
}.

\acknowledgements
Much of the efforts to assess the reliability of the results
were inspired by discussions with Bruce Lites.
He also provided the data employed here.
Thanks are due to Arturo L\'opez Ariste and Javier Trujillo Bueno for
suggestions to improve the manuscript.
This work has been partly funded by the Spanish DGES under project
95-0028-C. It has been carried out within
the EC-TMR European  Solar Magnetometry Network.
%

%
%-------- references section
%\bibliographystyle{/yavin/d/navarro/latex/bib/apj}
%\bibliography{/yavin/d/navarro/latex/bib/aamnem99,/yavin/d/navarro/latex/bib/articulos}
\newpage
%\bibliographystyle{apj}
%\bibliography{apjmnemonic,/home/jos/texto/papers/sun}

%
%------ captions
\newpage

%
%-figures with stokes profiles

%--

%--

%--
%--

%
% The very end
\end{document}